\documentclass{article}
\usepackage{spconf,amsmath,graphicx,color}
\usepackage{url,subfigure,cite,array}
\usepackage{hyperref} 
\usepackage{color}
\usepackage{booktabs}
\usepackage{tabularx}
\usepackage{multirow}

\title{Preserving background sound in noise-robust voice conversion via multi-task learning}
\name{
\begin{tabular}{c}
\it Jixun Yao$^1$, Yi Lei$^1$, Qing Wang$^1$, Pengcheng Guo$^1$, Ziqian Ning$^1$, Lei Xie$^1$,\\
\it Hai Li$^2$, Junhui Liu$^2$, Danming Xie$^2$
\end{tabular}
}
\address{$^1$Audio, Speech and Language Processing Group (ASLP@NPU)\\School of Computer Science, Northwestern Polytechnical University, Xi’an, China\\
$^2$iQIYI Inc, China}
\begin{document}
\ninept
\maketitle
\begin{abstract}
Background sound is an informative form of art that is helpful in providing a more immersive experience in real-application voice conversion (VC) scenarios. However, prior research about VC, mainly focusing on clean voices, pay rare attention to VC with background sound. The critical problem for preserving background sound in VC is inevitable speech distortion by the neural separation model and the cascade mismatch between the source separation model and the VC model. In this paper, we propose an end-to-end framework via multi-task learning which sequentially cascades a source separation (SS) module, a bottleneck feature extraction module and a VC module. Specifically, the source separation task explicitly considers critical phase information and confines the distortion caused by the imperfect separation process. The source separation task, the typical VC task and the unified task shares a uniform reconstruction loss constrained by joint training to reduce the mismatch between the SS and VC modules. Experimental results demonstrate that our proposed framework significantly outperforms the baseline systems while achieving comparable quality and speaker similarity to the VC models trained with clean data.

\end{abstract}
\section{Introduction}
\label{sec:intro}
Voice conversion (VC) is a speech signal transformation technique that changes the voice of the original speaker into the target speaker while keeping the linguistic content~\cite{sisman2020overview}. 
VC has attracted long-term research interests due to its various applications, such as voice dubbing for movies, personalized speech synthesis and speaker anonymization for voice privacy protection~\cite{mohammadi2017overview}.
With the increasing diversity of VC applications such as audiobook and movie dubbing, there has been rising need for handling the noisy source during conversion. First, the source speech in real applications is usually entangled with background sound, such as the mixed sound track of speech and background music in movies and audio books. This challenges the robustness of current VC system. Second, to provide better listening experiences, preservation of the original source background sound in the converted audio is also desired. It is ideal to have a controllable VC system that can keep or remove the source background sound in the target synthetic speech according to specific applications. Besides, background sound is also a valuable resource for VC-based data augmentation to improve the robustness of downstream systems, such as automatic speech recognition (ASR)~\cite{vcasrdata} or automatic speaker verification (ASV)~\cite{vcasvdata}.

Typical VC studies aim to disentangle the linguistic content and speaker timbre from source speech, without explicitly handling background sounds in the source speech~\cite{nvcnet,againvc,VQMIVC,robustvc}. As noisy source may affect the quality of the converted speech, some prior works~\cite{acvc,hqdu} addressed the background noise problem by using phonetic posteriorgrams (PPGs) or bottleneck features to represent the linguistic content, which is extracted from a multi-condition trained noise-robust automatic speech recognition (ASR) system. Another straightforward approach is to use an extra denoising model to handle the noisy source speech before voice conversion. Although noise is suppressed, inevitable speech distortion in the source speech induced by the imperfect denoising process will propagate to downstream and degrade the quality of the converted speech as well~\cite{sedis}.

For noise-controllable speech generation, adversarial training has been extensively studied in text-to-speech~\cite{lmxue,wnxu,jcong,syang}, while the generated speech can be clean or noisy conditioned on an acoustic tag no matter the speaker training data is clean or noisy. The problem is that the synthetic noisy speech only contains some kind of ``averaged" noise learned from the training data. Such an approach is not suitable for background sound preservation in the VC scenario. To address this problem, a noisy-to-noisy VC framework~\cite{n2n1,n2n} is proposed which adopts a pre-trained neural denoising module to conduct VC while preserving background sound at the same time. Specifically, the denoised speech is fed to the VC module and the residual (noisy speech subtracts the denoised speech) is regarded as the background sound which is finally combined with the converted speech. Similarly, helping with a per-trained source separation model, the singing voice conversion approach in~\cite{svc} manages to preserve background music (BGM) in the converted singing by simply concatenating the separated BGM with the converted speech.

The above approaches deal with noisy source by cascading individually trained denoising/separation model and VC model. Besides the different training objectives, as mentioned earlier, the inevitable distortion (error) in speech as well as in the background sound induced by the imperfect front-end module may lead to inferior sound quality with audible artifacts in the converted audio.

\begin{figure*}[ht]
        \centering
        \includegraphics[width=0.76\linewidth]{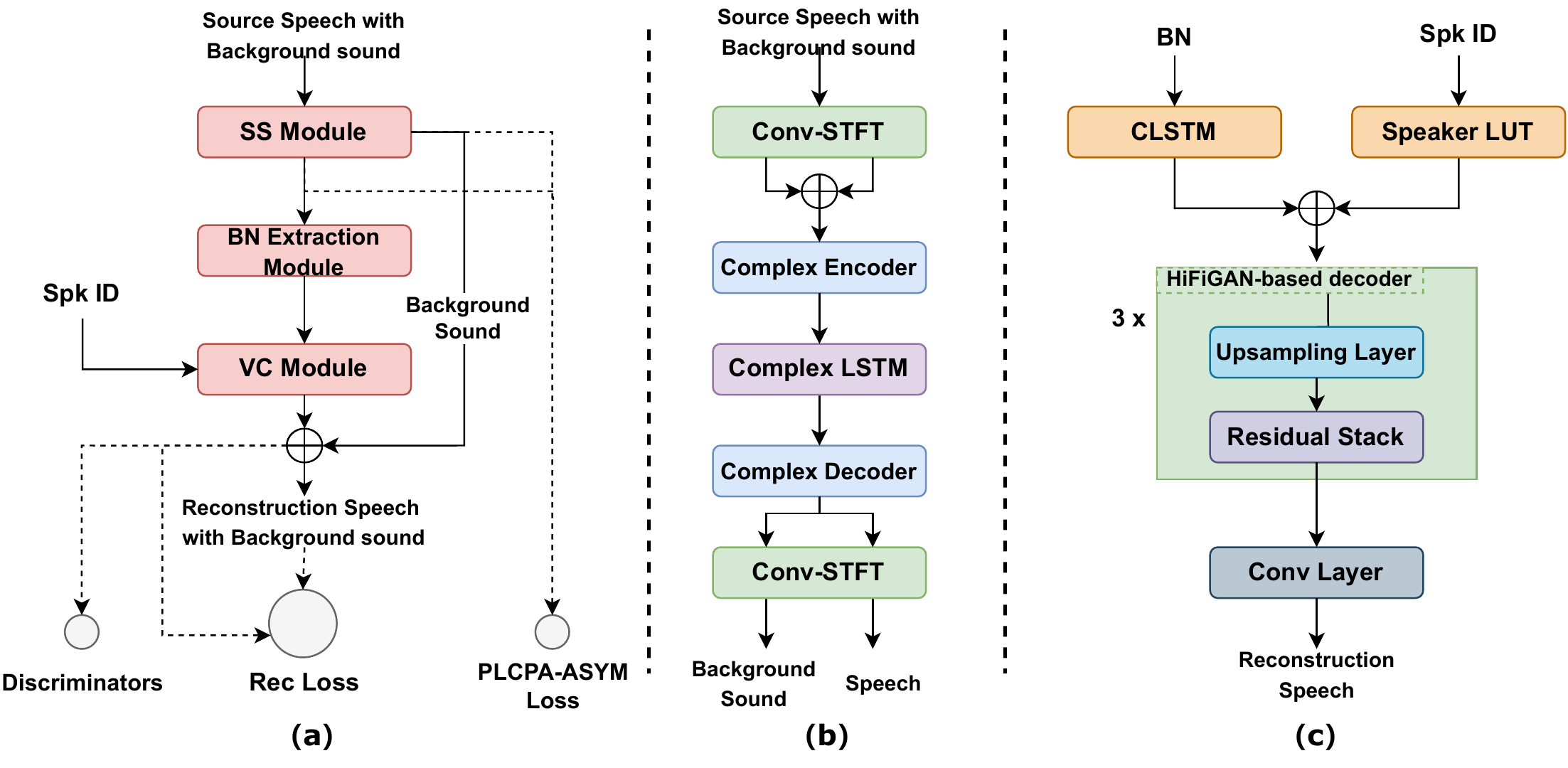}
        \vspace{-1em}
        \caption{(a) System overview of the proposed multi-task framework, BN represents bottleneck features. (b) SS module. (c) VC module. The solid line represents the forward propagation, and the dashed line represents which loss functions are used for the module output.}
         \vspace{-5pt}
        \label{fig:model}\vspace{-10pt}
\end{figure*}

To address this problem, we propose an end-to-end (E2E) framework in a multi-task learning manner for VC in the presence of background sound. We shape our E2E approach by sequentially cascading a source separation (SS) module, a bottleneck feature extraction module and a VC module with multiple specifically designed learning objectives. Specifically, the source separation task is formed with a deep complex convolution recurrent network (DCCRN)~\cite{dccrn} optimized by power-law compressed phase-aware (PLCPA) loss and asymmetry loss~\cite{phase}. Here critical phase information is explicitly considered and distortion is confined by leveraging the recent advances in neural source separation. Besides the source separation task and the typical VC task, importantly, the unified task shares a uniform reconstruction loss constrained by joint training to reduce the mismatch between the SS and VC modules while ensuring the quality of the converted speech. Experiments and ablation studies show the significant advantages of the proposed approach.

\section{Proposed Multi-Task Framework}
\label{sec:framework}


As shown in Figure~\ref{fig:model}(a), the proposed framework consists of three modules, e.g. the bottleneck feature extraction module, SS module, and VC module. With the above modules, we approach the problem of VC with background sound via a three-step process: 1) separating vocal and background sound from the input signal using the SS module, 2) conducting voice conversion on vocal using the VC module with the bottleneck feature as input, 3) superimposing the converted voice with background sound extracted by the SS module.
The bottleneck feature extraction module is a pre-trained ASR encoder developed by WeNet tools~\cite{yao2021wenet} and is available on the official website~\footnote{https://github.com/wenet-e2e/wenet}. The SS module and VC modules are trained beforehand with different datasets separately, and then jointly trained by multi-task learning with the bottleneck feature extraction module frozen.

\subsection{Source Separation Module}
Due to the large parameters and complex structure, most current source separation models are usually too bulky to be jointly trained with the VC task. Therefore, we adopt DCCRN as our SS module~\cite{dccrn}, which only has 3.7M parameters and achieved the best performance for Deep Noise Suppression (DNS) challenge in the real-time track~\cite{dns}. 

The original DCCRN estimates complex ratio mask (CRM) by the real and imaginary parts of the noisy complex spectrogram. 
In this paper, DCCRN is used as a source separation model by adding additional CRM of clean speech, as shown in Figure~\ref{fig:model} (b). Meanwhile, we use power-law compressed phase-aware asymmetric (PLCPA-ASYM) loss to replace the original scale-invariant signal-to-noise ratio (SI-SNR) loss to confine speech distortion and achieve better power control~\cite{phase}. 
Suppose $x$ and $\hat{x}$ are the estimated and clean spectrograms, respectively. PLCPA loss is defined as follows:

\begin{equation}
\begin{aligned}
\mathcal{L}_a(t, f) &=\left.|| x(t, f)\right|^p-\left.|\hat{x}(t, f)|^p\right|^2 \\
\mathcal{L}_p(t, f) &=\left.|| x(t, f)\right|^p e^{j \varphi(x(t, f))}-\left.|\hat{x}(t, f)|^p e^{j \varphi(\hat{x}(t, f))}\right|^2 \\
\mathcal{L}_{\texttt{plcpa}} &=\frac{1}{T} \frac{1}{F} \sum_t^T \sum_f^F\left(\alpha \mathcal{L}_a(t, f)+(1-\alpha) \mathcal{L}_p(t, f)\right),
\end{aligned}
\end{equation}
where $T$ and $F$ are the total time and frequency frames, respectively, while $t$ and $f$ stand for time and frequency index. The spectral compression factor $p$ is set to 0.3 and operator $\varphi$ calculates the argument of a complex number while the weighted coefficient between the phase-aware components and amplitude is $\alpha$.
The asymmetric loss~\cite{asymloss} is adapted to the amplitude part of the PLCPA loss to alleviate the over-suppression (OS) issue, which is defined as
\begin{equation}
\begin{aligned}
    h(x)&=\left\{  
             \begin{array}{lr}  
             0, \quad if\ \ x \leq 0, &  \\  
             x, \quad if\ \ x >0, &    
             \end{array}  \right. \\
    \mathcal{L}_{\texttt{os}}(t,f) &= \Big|h(|x(t,f)|^p - |\hat{x}(t,f)|^p)\Big|^2.\\
\end{aligned}
\end{equation}
So that the final PLCPA-ASYM loss can be defined as
\begin{equation}
    \mathcal{L}_{\texttt{plcpa-asym}} = \mathcal{L}_{\texttt{plcpa}} + \beta \frac{1}{T} \frac{1}{F} \sum_t^T \sum_f^F \mathcal{L}_{\texttt{os}}(t,f),
\end{equation}
where $\beta$ is the positive weighting coefficient for $\mathcal{L}_{\texttt{os}}(t,f)$.

\subsection{Voice Conversion Module}
After extracting the speech from source waveforms with background sound through the SS module, we hire a VC module to transform the speech into the target speaker. As illustrated in Figure~\ref{fig:model}(c), the VC module goes through single-stage training for efficient end-to-end learning.
The VC module consists of a convolutional long short-term memory (CLSTM) encoder and a HiFiGAN-based decoder~\cite{hifigan}, which aim at high-level linguistic representation encoding and waveform reconstruction, respectively. CLSTM consists of three stacks of convolution layers followed by the LeakyReLU activation function and a LSTM layer. 
The speaker embedding from the lookup table is fed to the decoder as conditions for target voice generation. 
The architecture and objective of the decoder generator follow the same configuration as HiFi-GAN~\cite{hifigan}.

\subsection{Multi-task Training}
To reduce the mismatch between the SS module and VC module during voice conversion in the presence of background sound, we adopt the multi-task learning strategy in the following steps: 1) only the VC module is optimized by the VC training loss with the SS module frozen; 2) only the SS module is optimized by PLCPA-ASYM loss with VC module frozen; 3) both the SS and VC modules are jointly optimized in a multi-task learning manner. The bottleneck feature extraction module is frozen at each stage.

\textbf{Reconstruction loss for the unified task.}
The unified task aims at optimizing the composed waveform with vocals and background sound by minimizing the reconstruction loss. 
We denote the Mel-spectrogram of training data as $M=(M_{\texttt{s}}, M_{\texttt{b}})$, which is the Mel-spectrogram composed of speech audio $X_{\texttt{s}}$ and background sound $X_{\texttt{b}}$. The reconstruction loss for the unified task is defined as
\begin{equation}
    \mathcal{L}_{\texttt{rec}}^{\texttt{uni}}=||M-\hat{M}||_{1},
\end{equation}
where $\hat{M}$ is the Mel-spectrogram of the reconstructed waveform containing speech voice and background sound.

\textbf{SS training loss.}
The source separation task considers critical phase information and uses PLCPA-ASYM loss to optimize the estimated CRM and confine the separated speech and background distortion.
PLCPA-ASYM loss is calculated for the separated speech and background sound, denoted by $\mathcal{L}_{\texttt{s}}^{\texttt{ss}}$ and $\mathcal{L}_{\texttt{b}}^{\texttt{ss}}$, respectively. 

\textbf{VC training loss.}
The objective of the typical VC task with no background sound follows HiFiGAN~\cite{hifigan}, which consists of reconstruction loss $\mathcal{L}_{\texttt{rec}}^{\texttt{vc}}$, feature matching loss $\mathcal{L}_{\texttt{fm}}^{\texttt{vc}}$, and adversarial loss $\mathcal{L}_{\texttt{adv}}^{\texttt{vc}}$. We adopt the L1 loss as reconstruction loss to optimize the spectrogram $\hat{M}_\texttt{s}$ of the predicted speech $\hat{X_\texttt{s}}$ as
\begin{equation}
    \mathcal{L}_{\texttt{rec}}^{\texttt{vc}}=||M_{\texttt{s}}-\hat{M_{\texttt{s}}}||_{1}.
\end{equation}
To improve the performance of voice conversion, we employ adversarial training for more natural speech. The adversarial generator loss of the single-task VC is calculated as
\begin{equation}
    \mathcal{L}_{\texttt{adv}}^{\texttt{gen}}=(D(\hat{X_{\texttt{s}}})-1)^2,
\end{equation}
\begin{equation}
    \mathcal{L}_{\texttt{adv}}^{\texttt{dis}}=(D(X_{\texttt{s}})-1)^2+D(\hat{X_{\texttt{s}}})^2,
\end{equation}
where $D$ is a discriminator network. For adversarial training stability, feature matching loss is also used as
\begin{equation}
    \mathcal{L}_{\texttt{fm}}^{\texttt{vc}}=\sum_{i=1}^T \frac{1}{N_i}\left\|D_i(X_{\texttt{s}})-D_i(\hat{X_{\texttt{s}}})\right\|_1,
\end{equation}
where $T$ denotes the total number of layers in the discriminator and $D_i$ produces the feature map of the $i$-th layer of the discriminator with $N_i$ number of features. 
The joint training process will combine the SS and VC modules into a single unit if the source separation losses and voice conversion losses are not added, rendering the separated intermediate results useless and unable to control whether the background sound is preserved.

The SS and VC modules share a uniform reconstruction loss constrained by joint training and the SS module can be regarded as data augmentation or a desirable pre-process model for the VC module to improve robustness. The final loss function of our multi-task learning approach is a weighted sum of the following losses:
\begin{equation}
\begin{aligned}
    \mathcal{L}_{\texttt{mtl}}=&\lambda_{\texttt{uni}}\mathcal{L}_{\texttt{rec}}^{\texttt{uni}}
    +\lambda_{\texttt{ss}}(\mathcal{L}_{\texttt{s}}^{\texttt{ss}}+\mathcal{L}_{\texttt{b}}^{\texttt{ss}})+\\
    &\lambda_{\texttt{vc}}(\mathcal{L}_{\texttt{rec}}^{\texttt{vc}}+\mathcal{L}_{\texttt{adv}}^{\texttt{vc}}+\mathcal{L}_{\texttt{fm}}^{\texttt{vc}}),
\end{aligned}
\end{equation}
where $\lambda_{\texttt{uni}}$, $\lambda_{\texttt{ss}}$, and $\lambda_{\texttt{vc}}$ are hyper-parameters balancing loss terms of each task. We set $\lambda_{\texttt{uni}}=45$, $\lambda_{\texttt{ss}}=1$ and $\lambda_{\texttt{vc}}=1$ empirically.

\section{Experiments}
\label{sec:framework}
\subsection{Dataset}
We train the SS module and VC module on two different datasets. To train the SS module, we use the MUSDB18-train dataset~\cite{musdb18}, which contains 100 full lengths of music tracks of different genres and their isolated drums, bass, vocals, and other items. The VC module is trained on VCTK dataset~\cite{veaux2016superseded} which contains 110 English speakers and 400 utterances per speaker. For the joint training of the SS and VC module, MUSDB18-train and VCTK datasets are mixed to conduct the evaluation of the background sound VC.
Background sound is randomly selected from the dataset and reprocessed to be clipped as the same length as speech data by a certain signal-to-noise ratio (SNR) range of $0$ to $10$ dB.
We randomly select 4 source speakers (i.e. p245, p264, p270, and p361) and 2 target speakers (i.e. p294 and p334) from the VCTK dataset mixed with clips from the MUSDB18-test dataset for conducting evaluations. For each source and target speaker, 30 utterances are reserved as the test set.
All training and evaluation data are conducted on 16 kHz sampling rate.

\subsection{Comparisons and Evaluation Metrics}
For a fair comparison, we conduct evaluations on the following systems with the same VC module: 1) Upper Bound, where the audio is converted from clean source voice and then superimposed with the original background sound; 2) Baseline1 (Sep), consisting of a VC module and a separation model~\footnote{https://github.com/bytedance/music\_source\_separation} by individual training, the same source separation approach as~\cite{svc}; 3) Baseline2 (Sep + Denoise), containing a VC module, the same separation model as Baseline1 and the same denoising model as~\cite{n2n}, which can be regarded as a more robust source separation way than~\cite{n2n}; 4) Proposed, the multi-task learning framework proposed in this paper.

The effectiveness of the proposed framework compared with the baseline systems is investigated by objective and subjective evaluations.
For subjective evaluation, we use comparative mean opinion score (CMOS) to compare the converted speech with background sound to the baseline system and use Upper Bound results as the reference to determine the ensemble quality. To evaluate the quality and similarity more effectively, we also conducted a mean opinion score (MOS) test on the converted speech without superimposing the background sound.
Scale-invariant-signal-to-distortion ratio (SI-SDR)~\cite{le2019sdr}, and perceptual evaluation of speech quality (PESQ)~\cite{pesq} are employed as the objective metrics to evaluate separated background sound and converted speech. Audio samples are available online~\footnote{\url{https://yaoxunji.github.io/background_sound_vc/}}.

\begin{table*}[h]\footnotesize
    \caption{Ensemble quality CMOS and speech MOS comparison of the proposed framework vs. different baseline systems. ``F" and ``M" represent female and male, corresponding to different gender combinations of source and reference voices.}\label{tab:mos}
    \centering
    \renewcommand\arraystretch{1.5}
    \renewcommand{\tabcolsep}{0.12cm}
    \vspace{3pt}
\begin{tabular}{c|c|ccccc|ccccc}
\hline
\multirow{2}{*}{System} & \multirow{2}{*}{CMOS} & \multicolumn{5}{c}{Quality} & \multicolumn{5}{c}{Similarity} \\ \cline{3-12} 
                  &                       & M2M            & M2F            & F2F            & F2M           & Mean           & M2M             & M2F            & F2F            & F2M            & Mean  \\ \hline
Upper Bound       & 0                     & 3.96$\pm0.07$  & 3.89$\pm0.08$  & 3.92$\pm0.07$  & 3.83$\pm0.08$ & 3.90$\pm0.08$  & 3.82$\pm0.06$   & 3.87$\pm0.06$  & 3.88$\pm0.07$  & 3.83$\pm0.07$  & 3.85$\pm0.07$   \\ \hline
Sep         & -0.54                 & 2.92$\pm0.10$  & 2.81$\pm0.11$  & 3.01$\pm0.11$  & 2.86$\pm0.11$ & 2.90$\pm0.11$  & 3.11$\pm0.10$   & 3.21$\pm0.09$  & 3.17$\pm0.08$  & 3.23$\pm0.10$  & 3.18$\pm0.09$   \\
Sep + Denoise         & -0.35                 & 3.44$\pm0.10$  & 3.35$\pm0.10$  & 3.49$\pm0.09$  & 3.40$\pm0.10$ & 3.42$\pm0.10$  & 3.40$\pm0.10$   & 3.44$\pm0.10$  & 3.45$\pm0.10$  & 3.47$\pm0.10$  & 3.44$\pm0.10$   \\
Proposed          & \textbf{-0.12}                 & \textbf{3.80$\pm$0.05}  & \textbf{3.69$\pm$0.06}  & \textbf{3.74$\pm$0.05}  & \textbf{3.65$\pm$0.06} & \textbf{3.72$\pm$0.05}  & \textbf{3.69$\pm$0.08}   & \textbf{3.74$\pm$0.06}  & \textbf{3.73$\pm$0.06}  & \textbf{3.64$\pm$0.07}  & \textbf{3.70$\pm$0.07}   \\
 \hline
\end{tabular}
\vspace{-15pt}
\end{table*}

\subsection{Experimental Results}
\subsubsection{Subjective Evaluation}
We first evaluate the overall performance of VC in the presence of background sound by CMOS tests between the compared systems and the Upper Bound, as shown in Table~\ref{tab:mos}. The results of CMOS, i.e. -0.54 of only separation approach, -0.35 of separation and denoising approach, and -0.12 of Proposed, demonstrate that the proposed framework achieves a statistically significant preference compared with baseline systems. In addition, the score of -0.12 indicates that the converted speech with background sound from the proposed framework has slight regression compared to Upper Bound in ensemble quality. 

To evaluate the quality and speaker similarity of the converted speech, we conduct MOS tests on different systems, as shown in Table\ref{tab:mos}. For more accurate rating results, we use the converted speech without superposing background sound for perceptive listening tests. We can find that the proposed framework achieves better results for the audio quality than the two baselines. Meanwhile, the MOS also indicates that the audio quality generated from in-gender voice conversion is better than cross-gender.
For the speaker similarity, the results show that our proposed framework significantly outperforms baselines in both cross-gender and in-gender.
The subjective results demonstrate the proposed multi-task learning framework can effectively deal with noise source and convert the voice to the target speaker with background sound.

\begin{table}[htb]
\vspace{-1.0em}
    \caption{The objective evaluation results of the converted speech and background sound before superimposing.}\label{tab:obj}
    \centering
    \renewcommand\arraystretch{1.1}
    \renewcommand{\tabcolsep}{0.38cm}
    \vspace{3pt}
\begin{tabular}{ccccc}
\hline
\multirow{2}{*}{System}          & \multicolumn{2}{c}{Speech}     & \multicolumn{2}{c}{Background Sound}        \\ \cline{2-5} 
          & SI-SDR         & PESQ          & SI-SDR         & PESQ          \\ \hline
Sep & 11.76          & 2.75          & 9.85           & 2.29          \\
Sep+Denoise & 12.01          & \textbf{3.43}  &9.85  & 2.29          \\
Proposed  & \textbf{12.10} & \textbf{3.43} & \textbf{11.11} & \textbf{2.56} \\ \hline
\end{tabular}
\vspace{-1em}
\end{table}

\subsubsection{Objective Evaluation}
For precise evaluations, we individually calculate the SI-SDR and PESQ results on speech and background sound before superimposing, since the superimposed audio may conceal the detail of speech and background sound. Because of the same separation model employed by model \textit{Sep} and \textit{Sep+Denoise}, the two baselines have the same SI-SDR and PESQ results of background sound. 

The objective results are shown in Table~\ref{tab:obj}. For evaluations on speech, the proposed framework achieves the highest scores on SI-SDR and PESQ than two baselines, which verifies the good performance of the proposed framework on the separated voice. In addition, we find that model Sep+Denoise obtains the same PESQ score with our proposed framework. It demonstrates our SS module can achieve similar separated speech results as the state-of-the-art denoising model which further indicates the effectiveness of the proposed SS module in multi-task learning.

Regarding background sound, our proposed framework outperforms two baselines and achieves the highest scores in the SI-SDR and PESQ metrics, i.e. 11.11 and 2.56, respectively. 
According to the objective results, our multi-task learning framework can achieve better separated results than baseline systems by a single source separation module.

\subsubsection{Ablation Study}
To further investigate the effectiveness of each component in our framework, we conduct ablation studies for speech and background sound by subjective and objective metrics, respectively, as shown in Table~\ref{tab:abl}. 
In the ablation studies, we remove the loss of the source separation task and the voice conversion task in the procedure of joint training. 

\begin{table}[htb]
\vspace{-0.8em}
    \caption{Ablation studies of speech MOS and background sound objective evaluation results. To evaluate the biases induced by each component, we remove them one at a time.}\label{tab:abl}
    \centering
    \renewcommand\arraystretch{1.2}
    \renewcommand{\tabcolsep}{0.13cm}
    \vspace{3pt}
\begin{tabular}{lcccc}
\hline
\multirow{2}{*}{Variants}                & \multicolumn{2}{c}{Speech}   & \multicolumn{2}{c}{Background Sound} \\ \cline{2-5} 
                                    & Quality       & Similarity      & SI-SDR      & PESQ      \\ \hline
Proposed                            & 3.69$\pm0.06$ & 3.71$\pm0.07$            & 11.11       & 2.56      \\
$\quad -\mathcal{L}_{*}^{\texttt{se}}$    & 3.36$\pm0.07$ & 3.51$\pm0.06$            & 7.91        & 1.96      \\
$\quad -\mathcal{L}_{*}^{\texttt{vc}}$    & 3.27$\pm0.07$ & 3.37$\pm0.07$            & 10.19       & 2.70      \\
$\quad -$Joint Training                   & 3.14$\pm0.05$ & 3.23$\pm0.06$            & 9.91        & 2.31     \\ \hline
\end{tabular}
\vspace{-1.0em}
\end{table}

The results demonstrate that variants without SS loss ``$-\mathcal{L}_{*}^{\texttt{ss}}$" and VC loss ``$-\mathcal{L}_{*}^{\texttt{vc}}$" get worse performance in both subjective and objective metrics. 
When removing ``$\mathcal{L}_{*}^{\texttt{ss}}$", the separated background sound degrades by a large margin in terms of SI-SDR and PESQ. Besides, the Mel-spectrogram of the vocal part is damaged severely. As for variant ``$-\mathcal{L}_{*}^{\texttt{vc}}$", it achieves worse speech quality and similarity than variant ``$-\mathcal{L}_{*}^{\texttt{ss}}$" but slightly better PESQ result than the proposed framework, which indicates that removing VC loss makes the variant pay more attention to separation.
In addition, the variant without joint training degrades speech and background sound performance.
We believe the reason is that there exists a mismatch between the SS module and VC module when using the separation results directly for conversion, leading to distortion of the conversion result.
The ablation studies verify the effectiveness of our proposed multi-task learning approach on VC with background sound.

\section{Conclusion}
\label{sec:conclusion}
In this paper, we present an end-to-end framework in a multi-task manner that can produce high-quality VC in the presence of background sound and is capable of flexible controlling linguistic content, background sound, and speaker timbre. The framework sequentially cascades a SS module, a bottleneck feature extraction module and a VC module. The SS module is formed with DCCRN and confines the distortion by leveraging the PLCPA-ASYM loss. To tackle the disentangling problem between the vocal and the background sound, we employ multi-task loss to constrain the outputs of the SS and VC modules. Furthermore, the source separation task, the voice conversion task, and the unified task share a uniform reconstruction loss to reduce the mismatch between the SS module and the VC module.
The subjective and objective evaluation results demonstrate that the proposed framework has higher speech quality and similarity, effectively bridging the margin on the ensemble quality between the baseline and the upper bound.

\bibliographystyle{IEEEbib}
\bibliography{strings,refs}

\begin{thebibliography}{10}

\bibitem{sisman2020overview}
Berrak Sisman, Junichi Yamagishi, Simon King, and Haizhou Li,
\newblock ``An overview of voice conversion and its challenges: From
  statistical modeling to deep learning,''
\newblock {\em IEEE/ACM Trans Audio Speech Lang. Process.}, vol. 29, pp.
  132--157, 2020.

\bibitem{mohammadi2017overview}
Seyed~Hamidreza Mohammadi and Alexander Kain,
\newblock ``An overview of voice conversion systems,''
\newblock {\em Speech Comm.}, vol. 88, pp. 65--82, 2017.

\bibitem{vcasrdata}
S.~Shahnawazuddin, Nagaraj Adiga, Kunal Kumar, Aayushi Poddar, and Waquar
  Ahmad,
\newblock ``Voice conversion based data augmentation to improve children's
  speech recognition in limited data scenario,''
\newblock in {\em Proc. Interspeech}, 2020, pp. 4382--4386.

\bibitem{vcasvdata}
S.~Shahnawazuddin, Waquar Ahmad, Nagaraj Adiga, and Avinash Kumar,
\newblock ``In-domain and out-of-domain data augmentation to improve children's
  speaker verification system in limited data scenario,''
\newblock in {\em Proc. ICASSP}, 2020, pp. 7554--7558.

\bibitem{nvcnet}
Bac Nguyen and Fabien Cardinaux,
\newblock ``Nvc-net: End-to-end adversarial voice conversion,''
\newblock in {\em Proc. ICASSP}, 2022, pp. 7012--7016.

\bibitem{againvc}
Yen{-}Hao Chen, Da{-}Yi Wu, Tsung{-}Han Wu, and Hung{-}yi Lee,
\newblock ``Again-vc: {A} one-shot voice conversion using activation guidance
  and adaptive instance normalization,''
\newblock in {\em Proc. ICASSP}, 2021, pp. 5954--5958.

\bibitem{VQMIVC}
Disong Wang, Liqun Deng, Yu~Ting Yeung, Xiao Chen, Xunying Liu, and Helen Meng,
\newblock ``{VQMIVC:} vector quantization and mutual information-based
  unsupervised speech representation disentanglement for one-shot voice
  conversion,''
\newblock in {\em Proc. Interspeech}, 2021, pp. 1344--1348.

\bibitem{robustvc}
Trung Dang, Dung~N. Tran, Peter Chin, and Kazuhito Koishida,
\newblock ``Training robust zero-shot voice conversion models with
  self-supervised features,''
\newblock in {\em Proc. ICASSP}, 2022, pp. 6557--6561.

\bibitem{acvc}
Damien Ronssin and Milos Cernak,
\newblock ``{AC-VC:} non-parallel low latency phonetic posteriorgrams based
  voice conversion,''
\newblock in {\em Proc. ASRU}, 2021, pp. 710--716.

\bibitem{hqdu}
Hongqiang Du, Lei Xie, and Haizhou Li,
\newblock ``Noise-robust voice conversion with domain adversarial training,''
\newblock {\em Neur. Net.}, vol. 148, pp. 74--84, 2022.

\bibitem{sedis}
Yangyang Xia, Sebastian Braun, Chandan K.~A. Reddy, Harishchandra Dubey, Ross
  Cutler, and Ivan Tashev,
\newblock ``Weighted speech distortion losses for neural-network-based
  real-time speech enhancement,''
\newblock in {\em Proc. ICASSP}, 2020, pp. 871--875.

\bibitem{lmxue}
Liumeng Xue, Shan Yang, Na~Hu, Dan Su, and Lei Xie,
\newblock ``Learning noise-independent speech representation for high-quality
  voice conversion for noisy target speakers,''
\newblock in {\em Proc. Interspeech}, 2022, pp. 2548--2552.

\bibitem{wnxu}
Wei{-}Ning Hsu, Yu~Zhang, Ron~J. Weiss, Yu{-}An Chung, Yuxuan Wang, Yonghui Wu,
  and James~R. Glass,
\newblock ``Disentangling correlated speaker and noise for speech synthesis via
  data augmentation and adversarial factorization,''
\newblock in {\em Proc. ICASSP}, 2019, pp. 5901--5905.

\bibitem{jcong}
Jian Cong, Shan Yang, Lei Xie, Guoqiao Yu, and Guanglu Wan,
\newblock ``Data efficient voice cloning from noisy samples with domain
  adversarial training,''
\newblock in {\em Proc. Interspeech}, 2020, pp. 811--815.

\bibitem{syang}
Shan Yang, Yuxuan Wang, and Lei Xie,
\newblock ``Adversarial feature learning and unsupervised clustering based
  speech synthesis for found data with acoustic and textual noise,''
\newblock {\em {IEEE} Signal Process. Lett.}, vol. 27, pp. 1730--1734, 2020.

\bibitem{n2n1}
Chao Xie, Yi{-}Chiao Wu, Patrick~Lumban Tobing, Wen{-}Chin Huang, and Tomoki
  Toda,
\newblock ``Noisy-to-noisy voice conversion framework with denoising model,''
\newblock in {\em Proc. APSIPA}, 2021, pp. 814--820.

\bibitem{n2n}
Chao Xie, Yi{-}Chiao Wu, Patrick~Lumban Tobing, Wen{-}Chin Huang, and Tomoki
  Toda,
\newblock ``Direct noisy speech modeling for noisy-to-noisy voice conversion,''
\newblock in {\em Proc. ICASSP}, 2022, pp. 6787--6791.

\bibitem{svc}
Divyesh~G. Rajpura, Jui Shah, Maitreya Patel, Harshit Malaviya, Kirtana
  Phatnani, and Hemant~A. Patil,
\newblock ``Effectiveness of transfer learning on singing voice conversion in
  the presence of background music,''
\newblock in {\em Proc. SPCOM}, 2020, pp. 1--5.

\bibitem{dccrn}
Yanxin Hu, Yun Liu, Shubo Lv, Mengtao Xing, Shimin Zhang, Yihui Fu, Jian Wu,
  Bihong Zhang, and Lei Xie,
\newblock ``{DCCRN:} deep complex convolution recurrent network for phase-aware
  speech enhancement,''
\newblock in {\em Proc. Interspeech}, 2020, pp. 2472--2476.

\bibitem{phase}
Sefik~Emre Eskimez, Takuya Yoshioka, Huaming Wang, Xiaofei Wang, Zhuo Chen, and
  Xuedong Huang,
\newblock ``Personalized speech enhancement: new models and comprehensive
  evaluation,''
\newblock in {\em Proc. ICASSP}, 2022, pp. 356--360.

\bibitem{yao2021wenet}
Zhuoyuan Yao, Di~Wu, Xiong Wang, Binbin Zhang, Fan Yu, Chao Yang, Zhendong
  Peng, Xiaoyu Chen, Lei Xie, and Xin Lei,
\newblock ``Wenet: Production oriented streaming and non-streaming end-to-end
  speech recognition toolkit,''
\newblock in {\em Proc. Interspeech}, 2021, pp. 4054--4058.

\bibitem{dns}
Chandan K.~A. Reddy, Harishchandra Dubey, Vishak Gopal, Ross Cutler, Sebastian
  Braun, Hannes Gamper, Robert Aichner, and Sriram Srinivasan,
\newblock ``{ICASSP} 2021 deep noise suppression challenge,''
\newblock in {\em Proc. ICASSP}, 2021, pp. 6623--6627.

\bibitem{asymloss}
Quan Wang, Ignacio Lopez{-}Moreno, Mert Saglam, Kevin~W. Wilson, Alan Chiao,
  Renjie Liu, Yanzhang He, Wei Li, Jason Pelecanos, Marily Nika, and Alexander
  Gruenstein,
\newblock ``Voicefilter-lite: Streaming targeted voice separation for on-device
  speech recognition,''
\newblock in {\em Proc. Interspeech}, 2020, pp. 2677--2681.

\bibitem{hifigan}
Jungil Kong, Jaehyeon Kim, and Jaekyoung Bae,
\newblock ``Hifi-gan: Generative adversarial networks for efficient and high
  fidelity speech synthesis,''
\newblock in {\em Proc. NeurIPS}, 2020.

\bibitem{musdb18}
Zafar Rafii, Antoine Liutkus, Fabian-Robert St{\"o}ter, Stylianos~Ioannis
  Mimilakis, and Rachel Bittner,
\newblock ``The {MUSDB18} corpus for music separation,'' 2017.

\bibitem{veaux2016superseded}
Christophe Veaux, Junichi Yamagishi, Kirsten MacDonald, et~al.,
\newblock ``Superseded-cstr vctk corpus: English multi-speaker corpus for cstr
  voice cloning toolkit,''
\newblock 2016.

\bibitem{le2019sdr}
Jonathan Le~Roux, Scott Wisdom, Hakan Erdogan, and John~R Hershey,
\newblock ``Sdr--half-baked or well done?,''
\newblock in {\em Proc. ICASSP}, 2019, pp. 626--630.

\bibitem{pesq}
Antony~W Rix, John~G Beerends, Michael~P Hollier, and Andries~P Hekstra,
\newblock ``Perceptual evaluation of speech quality (pesq)-a new method for
  speech quality assessment of telephone networks and codecs,''
\newblock in {\em Proc. ICASSP}, 2001, pp. 749--752.

\end{thebibliography}

\end{document}